\documentclass[aps,prd,preprint,groupedaddress]{revtex4-2}
\usepackage{bm}
\usepackage{amsmath}
\usepackage{amsfonts}
\usepackage{graphicx}
\usepackage{epstopdf}

\begin{document}
\title{Charged Particle Motion in Spherically Symmetric Distributions 
of Magnetic Monopoles}
\author{Robert Littlejohn}
\affiliation{Department of Physics, University of California, Berkeley}
\author{Philip Morrison}
\author{Jeffrey Heninger}
\affiliation{Department of Physics, University of Texas, Austin}
\email{robert@wigner.berkeley.edu}

\date{\today}

\begin{abstract}
The classical equations of motion of a charged particle in a
spherically symmetric distribution of magnetic monopoles can be
transformed into a system of linear equations, thereby providing a
type of integrability.   In the case of a single monopole, the
solution was given long ago by Poincar\'e.  In the case of a uniform
distribution of monopoles the solution can be expressed in terms of
parabolic cylinder functions (essentially the eigenfunctions of an
inverted harmonic oscillator).   This solution is relevant to recent
studies of nonassociative star products, symplectic lifts of
twisted Poisson structures and fluids and plasmas of electric and 
magnetic charges.
\end{abstract}
\maketitle

\newcommand{\Bvec}{\textbf{B}}
\newcommand{\Fvec}{\textbf{F}}
\newcommand{\rvec}{\textbf{r}}
\newcommand{\vvec}{\textbf{v}}
\newcommand{\avec}{\dot{\textbf{v}}}
\newcommand{\Avec}{\textbf{A}}
\newcommand{\pvec}{\textbf{p}}
\newcommand{\avechat}{{\hat{\textbf{a}}}}
\newcommand{\bvechat}{{\hat{\textbf{b}}}}
\newcommand{\fvechat}{{\hat{\textbf{f}}}}
\newcommand{\evechat}{{\hat{\textbf{e}}}}
\newcommand{\nvechat}{{\hat{\textbf{n}}}}
\newcommand{\xvechat}{{\hat{\textbf{x}}}}
\newcommand{\yvechat}{{\hat{\textbf{y}}}}
\newcommand{\zvechat}{{\hat{\textbf{z}}}}
\newcommand{\omegavec}{\bm{\omega}}
\newcommand{\sigmavec}{\bm{\sigma}}
\newcommand{\Rdot}{{\dot R}}
\newcommand{\tr}{{\mathop{\textrm{tr}}}}
\newcommand{\udot}{{\dot u}}
\newcommand{\ubar}{{\bar u}}
\newcommand{\wbar}{{\bar w}}
\newcommand{\rdot}{{\dot r}}
\newcommand{\Abar}{{\bar A}}
\newcommand{\Bbar}{{\bar B}}
\newcommand{\sbar}{{\bar s}}
\newcommand{\Reals}{\mathbb{R}}
\newcommand{\Complexes}{\mathbb{C}}

\section{\label{intro}Introduction}

Recent studies of nonassociative star products by Bakas and L\"ust
\cite{BL} have led to models involving the classical motion of charged
particles in distributions of magnetic monopoles.  Such models are
nonassociative because the obvious definition of the Poisson bracket
fails to satisfy the Jacobi identity\cite{PJM82,HM20}.  In these
models this failure can be expressed in terms of a closed 3-form,
essentially $\nabla\cdot\Bvec$, which qualifies the structure as
twisted Poisson.  Such systems pose a challenge to quantization, since
all the usual methods depend in one way or another on the Jacobi
identity.  Star quantization of monopole systems has been of interest
for some time, see for example Cari\~nena et al \cite{Cetal} and
Soloviev \cite{Sol}, but the use of nonassociative star products for
distributions of monopoles is more recent.

In a different approach Kupriyanov and Szabo \cite{KS} have tackled
such systems directly and provided a symplectic lift in a phase space
of doubled dimensionality, that is, in a fully associative framework
in which the usual Jacobi identity is valid.  In addition, some work
has been done on the continuum Poisson structures associated with
fluids and plasmas containing distributions of monopoles.  It turns
out that in most interesting circumstances these (continuum) Poisson
structures are also nonassociative \cite{HM20}.  In fact, Lainz, Sard\'on
and Weinstein \cite{LSW} have shown that twisted Poisson
structures for particle motion correspond to fluid structures that are
not only not Poisson, they are not even twisted Poisson. We ourselves
have had a long-standing interest in Poisson structures in fluids and
plasmas \cite{PJM82,Mo80,WM81,MW82} and in charged particle motion in
magnetic fields \cite{Li83}.

The case of spherically symmetric distributions of monopoles is
especially interesting, because the magnetic field is rotationally
invariant and in a symplectic setting this would be enough to produce
a conserved angular momentum that would lead to complete
integrability.  Indeed, if the distribution consists of a single
monopole, this is exactly what happens; the problem of a single
charged particle moving in the field of a single monopole was first
solved by Poincar\'e \cite{Po1896}, who exhibited the conserved angular
momentum and showed that the motion lies on a cone whose apex is the
monopole.  Particle motion in other spherically symmetric
distributions of monopoles, including the uniform distribution, has
resisted complete solution.  Bakas and L\"ust \cite{BL} produced some
first integrals and derived some quantitative constraints on the
motion, but did not obtain a complete solution; and Kupriyanov and
Szabo \cite{KS}, with their doubled symplectic structure, found some
integrals in involution, but not enough to produce complete
integrability (they needed twice the usual number, because of their
doubled symplectic structure, and, in particular, they did not find an
angular momentum vector).

In this article we will exhibit a transformation that makes the
equations of motion of a charged particle in the field of a
spherically symmetric distribution of magnetic monopoles a linear
system.  In the case of a uniform distribution of monopoles we will
show that the solution can be given in terms of parabolic cylinder
functions, that is, essentially the energy eigenfunctions of an
inverted harmonic oscillator.  In addition we find an $S$-matrix
connecting asymptotic states as $t\to\pm\infty$.  We do not call this
complete integrability because the usual definition of integrability
\cite{Ar78} requires a symplectic structure, but the system is
completely integrable in most ordinary senses.

\section{The Solution}

\subsection{The Setup and Some First Integrals}

Before specializing to the spherically symmetric case, we note that
the general nonrelativistic equation of motion for a particle of mass
$m$ and electric charge $e$ in any magnetic field $\Bvec$ is assumed
to be
\begin{equation}
\ddot{\rvec}=\frac{e}{mc} \vvec\times\Bvec,
  \label{geneom}
  \end{equation}
regardless of whether $\nabla\cdot\Bvec=0$ or not.  Here
$\vvec=\dot{\rvec}$ is the velocity and $\ddot{\rvec}=\dot{\vvec}$ is
the acceleration of the particle.  In the symplectic setting, that is,
when $\nabla\cdot\Bvec=0$, the equations of motion (\ref{geneom})
preserve the volume form $d^3\rvec\,d^3\pvec$ in phase space, according
to the usual Liouville theorem, and moreover this form is proportional
to $d^3\rvec\,d^3\vvec$.  In the nonsymplectic setting
($\nabla\cdot\Bvec\ne0)$ the vector potential and hence the canonical
momentum $\pvec=m\vvec-(e/c)\Avec$ are not defined.  The form
$d^3\rvec\,d^3\vvec$, however, is defined and is preserved by the flow,
as shown by the calculation
 \begin{equation}
 \frac{\partial} {\partial \rvec} \cdot \dot\rvec +  
 \frac{\partial} {\partial \vvec} \cdot \dot\vvec=0.
 \end{equation}
Thus, the system possesses a preserved volume form in phase space even
in the nonsymplectic setting.

We now specialize to a spherically symmetric distribution of magnetic
monopoles with density $\rho(r)$, so that Maxwell's equation is
$\nabla\cdot\Bvec=4\pi\rho(r)$.  We denote the magnetic charge inside
radius $r$ by $g(r)$,
\begin{equation}
  g(r)=4\pi\int_0^r r^{\prime2}\,dr'\,\rho(r'),
\end{equation}
so that Gauss's law gives
\begin{equation}
  \Bvec(\rvec) = g(r)\frac{\rvec}{r^3}.
  \end{equation}
Then the equation of motion in this magnetic field is
\begin{equation}
\ddot{\rvec}=h(r)\vvec\times\rvec,
  \label{theeom}
  \end{equation}
with  
\begin{equation}
  h(r)=\frac{e}{mc}\frac{g(r)}{r^3}.
  \end{equation}
Parts of the solution to this system have been given by Bakas and
L\"ust \cite{BL}.

It follows from (\ref{theeom}) that
$\rvec\cdot\avec=\vvec\cdot\avec=0$, which implies 
\begin{equation}
  \frac{dv^2}{dt} = 2\vvec\cdot\avec=0,
  \label{Econserv}
  \end{equation}
or
\begin{equation} 
  \vvec^2=v^2=v_0^2=\text{const}.
  \label{v0eqn}
  \end{equation}
This is the first integral of
conservation of energy, where the 0-subscript indicates initial
conditions at $t=0$.   

As a first special case, we take $\vvec_0=0$, which by
(\ref{v0eqn}) implies $\vvec=0$ and $\rvec=\text{const.}$ for all
$t$.  We henceforth dispense with this case by assuming $\vvec_0\ne0$,
which implies that $\vvec\ne0$ for all $t$.

We obtain a second first integral by noting  that
\begin{equation}
  \frac{d}{dt}|\vvec\times\rvec|^2 = 2(\avec\times\rvec)\cdot
  (\vvec\times\rvec)= 2\left|\left|
  \begin{array}{cc}
    \avec\cdot\vvec & \avec\cdot\rvec \\
    \rvec\cdot\vvec & \rvec\cdot\rvec
    \end{array}\right|\right|=0,
  \end{equation}
so that $|\vvec\times\rvec|^2=|\vvec_0\times\rvec_0|^2$.   This leads
to a second special case, in which $\vvec_0\times\rvec_0=0$, which
implies that $\vvec\times\rvec=0$ for all $t$.  This in turn implies
that $\rvec$ and $\vvec$ are parallel with $\avec=0$, so that
$\rvec(t)=\vvec_0t+\text{const}$, and the particle moves on a radial
line with constant velocity.   We henceforth dispense with this
special case by assuming that $\vvec_0\times\rvec_0\ne0$, which
implies that $\vvec\times\rvec\ne0$ for all $t$.

Since $(d/dt)r^2=2\rvec\cdot\vvec$ and
$(d/dt)(\rvec\cdot\vvec)=\vvec^2=v_0^2$, we have
$\rvec\cdot\vvec=\rvec_0\cdot\vvec_0+v_0^2t$, and we see that there exists a
unique time $t$ such that $\rvec\cdot\vvec=0$.  We henceforth take
this time to be $t=0$, so that $\rvec_0\cdot\vvec_0=0$ and 
\begin{equation}
  \rvec\cdot\vvec=v_0^2t.
  \label{vdotreqn}
  \end{equation}
Using $(d/dt)r^2=2\rvec\cdot\vvec$ together with \eqref{vdotreqn}  implies 
\begin{equation}
  r^2=r_0^2+v_0^2t^2,
  \label{rsqeqn}
  \end{equation}
and we see that the particle reaches a minimum distance from the
origin $r_0$ at $t=0$.   This minimum distance cannot be zero, since
we are assuming that $\rvec_0\times\vvec_0\ne0$.  Also note that with
this choice of initial time,
\begin{equation}
  |\vvec\times\rvec|^2 = v_0^2r_0^2.
  \label{vcrossreqn}
  \end{equation}

\subsection{The Frame}

We also note that
\begin{equation}
  |\vvec\times(\vvec\times\rvec)|^2 = v^2|\vvec\times\rvec|^2 =
  v_0^2 |\vvec_0\times\rvec_0|^2 = v_0^4r_0^2=\text{const},
  \end{equation}
and that under our assumptions this constant is nonzero. Thus we
have three nonvanishing, mutually orthogonal vectors,
 \begin{equation}
  \Fvec_1=\vvec\times(\vvec\times\rvec), \qquad
  \Fvec_2=\vvec\times\rvec, \qquad
  \Fvec_3=-\vvec,
  \label{Fvecdef}
  \end{equation}
whose magnitudes are constants.  These vectors must evolve by means
of a time-dependent rotation, much as in rigid body theory.

To find this rotation we differentiate the three vectors with respect
to time.  First, $d\vvec/dt$ is given by (\ref{theeom}).   Next, we
have
\begin{equation}
  \frac{d}{dt}(\vvec\times\rvec)=\avec\times\rvec=
  h(r)[(\rvec\cdot\vvec)\rvec-r^2\vvec].
  \end{equation}
We use the identity
\begin{equation}
  \rvec=\frac{1}{v^2}[\vvec(\vvec\cdot\rvec)-\vvec\times(\vvec\times\rvec)]
  \label{aux}
  \end{equation}
in this to express the result as a linear combination of the three
$\Fvec$'s, obtaining
\begin{equation}
  \frac{d}{dt}(\vvec\times\rvec)= -\frac{h(r)}{v^2}
  [(\vvec\times\rvec)^2\,\vvec + (\vvec\cdot\rvec)\,\vvec\times
  (\vvec\times\rvec)],
  \label{vrdot}
  \end{equation}
or, with the help of (\ref{v0eqn}), (\ref{vdotreqn}) and 
(\ref{vcrossreqn}),
\begin{equation}
  \frac{d}{dt}(\vvec\times\rvec)=-h(r)[r_0^2\,\vvec+
  t\vvec\times(\vvec\times\rvec)].
  \end{equation}
As for the third vector, we have
\begin{equation}
  \frac{d}{dt}[\vvec\times(\vvec\times\rvec)] = 
  \vvec\times\frac{d}{dt}(\vvec\times\rvec)=
  -h(r)t\vvec\times[\vvec\times(\vvec\times\rvec)]=
  h(r)tv_0^2\,\vvec\times\rvec.
  \label{vvrdot}
  \end{equation}

We summarize (\ref{theeom}), (\ref{vrdot}) and (\ref{vvrdot}) by writing
\begin{equation}
  \frac{d}{dt}\left(\begin{array}{ccc}
      \Fvec_1 & \Fvec_2 & \Fvec_3
      \end{array}\right) = 
	\left(\begin{array}{ccc}
      \Fvec_1 & \Fvec_2 & \Fvec_3
      \end{array}\right)
	h(r)
    \left(\begin{array}{ccc}
        0 & -t & 0\\
        tv_0^2 & 0 & -1\\
        0 & r_0^2 & 0
        \end{array}\right).
    \label{derivarray}
    \end{equation}
We now define orthonormal unit vectors of a ``body frame,''
\begin{equation}
  \fvechat_1=\frac{\Fvec_1}{v_0^2r_0}, \qquad
  \fvechat_2=\frac{\Fvec_2}{v_0r_0}, \qquad
  \fvechat_3=\frac{\Fvec_3}{v_0}.
  \label{fvecdef}
  \end{equation}
This frame is right-handed,
\begin{equation}
  \fvechat_i \times \fvechat_j = \epsilon_{ijk}\,\fvechat_k,
  \label{fcrossf}
  \end{equation}
where we use the summation convention.  In terms of these vectors
(\ref{derivarray}) becomes
\begin{equation}
  \frac{d}{dt}\left(\begin{array}{ccc}
      \fvechat_1 & \fvechat_2 & \fvechat_3
      \end{array}\right) =
	\left(\begin{array}{ccc}
          \fvechat_1 & \fvechat_2 & \fvechat_3
      \end{array}\right) 
	h(r)
    \left(\begin{array}{ccc}
        0 & -v_0t & 0\\
        v_0t & 0 & -r_0\\
        0 & r_0 & 0
        \end{array}\right).
    \label{bderivarray}
    \end{equation}
We write this equivalently as
\begin{equation}
  \frac{d\fvechat_i}{dt} = \fvechat_j \, \Omega_{ji},
  \label{dfdt}
    \end{equation}
where the matrix $\Omega$ is defined by (\ref{bderivarray}) (including
the factor $h(r)$).

Note that since $r$ is a known function of time [see (\ref{rsqeqn})],
$\Omega$ is too.  Thus, (\ref{bderivarray}) or (\ref{dfdt}) is a
system of linear differential equations with time-dependent
coefficients. Given its solution one can use \eqref{aux}, with
\eqref{v0eqn} and \eqref {vdotreqn}, to write our solution of
\eqref{theeom} as
\begin{equation}
  \rvec=  -v_0t \,\fvechat_3 - r_0\,\fvechat_1
  \,.
  \label{sol}
\end{equation}

Equation~\eqref{dfdt} defines the Frenet-Serret apparatus for a space
curve with $s=v_0t$ being arc length. Because $\fvechat_3$ is
anti-tangent to the curve, the curvature is $\kappa(s)= h(r) r_0/v_0$
and $\fvechat_2$ is the unit normal. Similarly, $\fvechat_1$ is the
binormal and the torsion $\tau(s)=-h(r)t=-h(r)s/v_0$. So
$\tau/\kappa=-v_0t/r_0=-s/r_0$, from which it follows by an
observation due to Enneper that the orbits are geodesics on a cone
(see \cite{Blaschke} p.\ 47).  This generalizes Poincar\'e's result
for a single monopole \cite{Po1896}.  Rather than go through the
details of Enneper's analysis we will provide a simpler and more
direct derivation of these conclusions in Sec.~\ref{conesgeodesics}.

We will now put (\ref{dfdt}) into a more convenient form with some
ideas from rigid body theory.  We define a ``space frame''
$\evechat_i$ as the unit vectors of the coordinate axes (which are
time-independent), and we define the angular velocity $\omegavec$ of
the body frame relative to the space frame by
\begin{equation}
  \frac{d\fvechat_i}{dt}=\omegavec\times\fvechat_i.
  \end{equation}
Now writing $\omegavec=\omega_k\fvechat_k$ and using (\ref{fcrossf}),
we obtain $d\fvechat_i/dt = \omega_k \,\epsilon_{kij}\,\fvechat_j$,
which, combined with (\ref{dfdt}), gives
\begin{equation}
  \Omega_{ij} =-\epsilon_{ijk}\,\omega_k, \qquad
  \omega_i = -\frac{1}{2}\epsilon_{ijk}\,\Omega_{jk},
  \end{equation}
the usual isomorphism between $\mathfrak{so}(3)$ and $\Reals^3$.
Finally, comparing this with (\ref{bderivarray}) we find
\begin{equation}
  \omegavec=h(r)(r_0\,\fvechat_1 + v_0t\,\fvechat_3).
  \label{omegaeqn}
  \end{equation}

Now by combining (\ref{sol}) and (\ref{omegaeqn}) we see that
\begin{equation}
  \omegavec=-h(r)\rvec =-\frac{e\Bvec(r)}{mc},
  \label{darboux}
  \end{equation}
so the instantaneous axis of rotation of the body frame, apart from
sign, is in the direction of the magnetic field, that is, in the
radial direction, while its magnitude is the instantaneous
gyrofrequency at the particle position.  This would obviously be true
in a uniform magnetic field but the fact that it generalizes to the
nonuniform fields considered here was not obvious to us.

Now we break the velocity of the particle into its parallel and
perpendicular components along the radial direction, so that $v^2 =
v_\parallel^2+v_\perp^2$.  We note that (\ref{vdotreqn}) implies
$rv_\parallel=v_0^2t$, so that
\begin{equation}
  v_\perp = \sqrt{v_0^2-v_\parallel^2} = \frac{v_0r_0}{r},
  \label{vperpeqn}
  \end{equation}
where we use (\ref{rsqeqn}).  This gives us the effective gyroradius,
\begin{equation}
  r_g = \frac{v_\perp}{|\omegavec|} = \frac{v_0r_0}{r^2h(r)}.
  \label{rgeqn}
  \end{equation}

Now let us define a rotation operator $R$ by $\fvechat_i=R\evechat_i$,
that is, $R$ maps the space frame into the body frame.  We write the
matrix elements of $R$ in the space frame as $R_{ij} =
\evechat_i\cdot(R\evechat_j)=\evechat_i\cdot\fvechat_j$.  Then by
inserting a resolution of the identity we have
\begin{equation}
   \frac{d\fvechat_i}{dt} = \Rdot\evechat_i=
	\fvechat_j [\fvechat_j\cdot(\Rdot\evechat_i)]=
	\fvechat_j [(R\evechat_j)\cdot(\Rdot\evechat_i)]=
	\fvechat_j [\evechat_j\cdot(R^{-1}\Rdot\evechat_i)],
        \label{fvechatdef}
	\end{equation}
so that $\Omega=R^{-1}\Rdot$, or
\begin{equation}
   \Rdot = R\Omega,
   \label{Reqn}
   \end{equation}
where we now write $R$ for the matrix with components $R_{ij}$.

This is a convenient form for the linearized equations; the solution
is a curve $R(t)\in SO(3)$.  The matrix $\Omega$ belongs to the Lie
algebra, a convenient basis for which is the set of matrices $J_i$,
defined by $(J_i)_{jk} = -\epsilon_{ijk}$.  These satisfy
\begin{equation}
  [J_i,J_j] = \epsilon_{ijk}\, J_k, \label{Liealgebra} \end{equation}
and we have
\begin{equation}
   \Omega=h(r)(r_0\, J_1 + v_0t\,J_3),
   \label{Omegaeqn}
   \end{equation}
which has the same content as (\ref{omegaeqn}).

\subsection{The Lift Into $SU(2)$}

To solve (\ref{Reqn}) with $\Omega$ given by (\ref{Omegaeqn}) we lift
the curve $R(t)\in SO(3)$ to a curve $u(t) \in SU(2)$.   The
projection $:SU(2)\to SO(3)$ is
\begin{equation}
  R_{ij} = \frac{1}{2} \tr(u^\dagger \sigma_i u \sigma_j),
  \label{uRproj}
  \end{equation}
where $u\in SU(2)$ and $\sigma_i$ are the Pauli matrices.  
The map $:u\mapsto R(u)$ is a homomorphism. Also, the
map $J_i \mapsto -(i/2)\sigma_i$ is the isomorphism between the
Lie algebras.  Thus, the lifted equation of motion is
\begin{equation}
   \udot = -\frac{i}{2}h(r)\, u(r_0\, \sigma_1 + v_0t\,\sigma_3).
   \end{equation}

We call the solution $u(t)$ that satisfies $u(0)=1$ ``the fundamental
solution,'' where $1$ is the identity matrix.  It suffices to
determine this solution, since any other that satisfies $u(0)\ne1$ is
given by $u(t)=u(0)u_f(t)$, where $u_f(t)$ is the fundamental
solution.  In the fundamental solution the space and body frames
coincide at $t=0$.

It is slightly more convenient to work with $w=u^\dagger$, a
substitution that converts right-actions into left-actions.  Also, we
note that the second column of an element of $SU(2)$ is the
time-reversed image of the first column, that is,
$w_{12}=-\wbar_{21}$, $w_{22}=\wbar_{11}$, where the overbar indicates
complex conjugation, so it suffices to solve for the first column of
$w$.  Thus the equations of motion become
\begin{equation}
  \frac{d}{dt}\left(\begin{array}{c}
      w_{11} \\ w_{21}
      \end{array}\right)=
      \frac{i}{2}h(r) \left(\begin{array}{cc}
        v_0t & r_0 \\
        r_0 & -v_0t
        \end{array}\right)
      \left(\begin{array}{c}
        w_{11} \\ w_{21}
        \end{array}\right).
      \label{weqn}
    \end{equation}

\subsection{Uniform Sphere of Monopolium}

We now specialize to the case $\rho(r)=\rho_0=\text{const.}$, which
implies $g(r)=(4\pi/3)\rho_0\,r^3$ and $h(r)=(4\pi/3)(\rho_0
e/mc)=h_0=\text{const}$.  In the following we assume $h_0>0$; the case
$h_0<0$ is handled similarly.  The problem has one dimensionless
parameter, which we denote by
\begin{equation}
   p=\sqrt{h_0r_0^2/4v_0}.
\end{equation}
An interpretation of this parameter is given below.

It is interesting that the equations of motion (\ref{weqn}) now become
the normal form for Landau-Zener transitions in one dimension
\cite{FL,YCdV,TK}.  Taking the second derivative causes these equations 
to decouple, and we obtain,
\begin{subequations}
\label{weqns1}
\begin{eqnarray}
  \frac{d^2w_{11}}{dt^2}+\left[\frac{h_0^2}{4}(r_0^2+v_0^2t^2)
    -\frac{ih_0v_0}{2}\right]w_{11}&=0,
  \label{weqns1a}\\
  \frac{d^2w_{21}}{dt^2}+\left[\frac{h_0^2}{4}(r_0^2+v_0^2t^2)
    +\frac{ih_0v_0}{2}\right]w_{21}&=0.
  \label{weqns1b}
  \end{eqnarray}
\end{subequations}
The solution can be expressed in terms of parabolic cylinder
functions, essentially the energy eigenfunctions for an inverted
harmonic oscillator.  The properties of these functions that are
needed for this paper have been summarized in Appendix~\ref{pcfuns}.

By the substitution 
\begin{equation}
  z=e^{-i\pi/4}\omega_0 t
  \label{zdef}
  \end{equation}
where $\omega_0=\sqrt{h_0v_0}$ the equations (\ref{weqns1}) become
\begin{subequations}
\label{weqns2}
\begin{eqnarray}
  \frac{d^2w_{11}}{dz^2}+\left(ip^2+\frac{1}{2}
    -\frac{z^2}{4}\right)w_{11}&=0,
  \label{weqns2a}\\
  \frac{d^2w_{21}}{dz^2}+\left(ip^2-\frac{1}{2}
    -\frac{z^2}{4}\right)w_{21}&=0.
  \label{weqns2b}
  \end{eqnarray}
\end{subequations}
According to (\ref{pcode}) the solutions can be written,
\begin{subequations}
  \label{wDsolns}
\begin{eqnarray}
  w_{11}(t)&=&A_1 D_\nu(z) + B_1 D_\nu(-z),\\
  w_{21}(t)&=&A_2 D_{\nu-1}(z) + B_2 D_{\nu-1}(-z),
  \end{eqnarray}
\end{subequations}
where $D_\nu$ is a parabolic cylinder function, where $A_{1,2}$ and
$B_{1,2}$ are constants and where
\begin{equation}
	\nu=ip^2.
	\label{nudef}
	\end{equation}

The solutions (\ref{wDsolns}) must satisfy (\ref{weqn}) with
$h(r)=h_0$.   By the substitution (\ref{zdef}) the latter equations
can be written
\begin{equation}
  \frac{d}{dz}\left(\begin{array}{c}
      w_{11} \\ w_{21}
      \end{array}\right)=
      \left(\begin{array}{cc}
       -z/2 & -e^{-i\pi/4}p\\
        -e^{-i\pi/4}p & z/2
        \end{array}\right)
      \left(\begin{array}{c}
        w_{11} \\ w_{21}
        \end{array}\right).
      \label{weqnz}
    \end{equation}
Now substituting (\ref{wDsolns}) into these and using the recursion
relations (\ref{recursionrels}) we find that the four constants are related by
\begin{equation}
  A_2=pe^{-i\pi/4}A_1,\qquad B_2=-pe^{-i\pi/4}B_1.
  \end{equation}
Finally, by imposing the initial conditions $w_{11}(0)=1$,
$w_{21}(0)=0$ of the fundamental solution, we find the first column of
$w$ (and the second row of $u$) in the form

\begin{subequations}
\label{wcol1}
\begin{eqnarray}
  w_{11}(t)&=&u_{22}(t)=A[D_\nu(z) + D_\nu(-z)],\\
  w_{21}(t)&=&-u_{21}(t)=Ape^{-i\pi/4}\,[D_{\nu-1}(z) 
	- D_{\nu-1}(-z)],
  \end{eqnarray}
\end{subequations}
where
\begin{equation}
  A=\frac{\Gamma\bigl((1-\nu)/2\bigr)}{2^{(\nu+1)/2}\sqrt{2\pi}}=
  \frac{1}{2D_\nu(0)}.
  \end{equation}
Then by using $w_{22}=\wbar_{11}$, $w_{12}=-\wbar_{21}$, we find the
second column of $w$ (and the first row of $u$),
\begin{subequations}
\label{wcol2}
\begin{eqnarray}
  w_{12}(t)&=&-u_{12}(t)=\frac{B}{pe^{-i\pi/4}}
	[D_\nu(z)-D_\nu(-z)],\\
  w_{22}(t)&=&u_{11}(t)=B[D_{\nu-1}(z)+D_{\nu-1}(-z)],
  \end{eqnarray}
\end{subequations}
where
\begin{equation}
  B=\frac{\Gamma(1-\nu/2)}{2^{\nu/2}\sqrt{2\pi}}=
	\frac{1}{2D_{\nu-1}(0)}.
  \end{equation}
The constants $A$ and $B$ satisfy the identities,
\begin{equation}
  \frac{1}{|A|^2} \pm \frac{p^2}{|B|^2}=4e^{\pm p^2\pi/2},
  \qquad AB = \frac{\Gamma(1-\nu)}{2\sqrt{2\pi}}.
  \label{ABident}
  \end{equation}
In taking the complex conjugates of $w_{11}$ and $w_{21}$ it helps to
notice that $\overline{D_\nu(z)}=D_{\bar\nu}({\bar z})=D_{-\nu}(iz)$,
and, one must work with the linear dependencies (\ref{lindeps})
satisfied by the functions $D_\nu$ to bring the results into the form
shown.  One must also use various identities satisfied by the
$\Gamma$-function.  To check the results we can show that $\det w=1$ for all
$t$; in this calculation $\det w$ turns out to be proportional to the
Wronskian of the solutions $D_\nu(\pm z)$.

The matrix $u(t)\in SU(2)$, given by (\ref{wcol1}) and (\ref{wcol2}),
can be projected onto the matrix $R(t) \in SO(3)$ via (\ref{uRproj}).
Actually, it is easier to write $u$ in axis-angle form, and then to
use the same axis and angle for $R$.  The matrix $R(t)$, so
determined, is the fundamental solution; from it the time evolution of
the vectors $\fvechat_i$ follows from the definition of $R$, that is,
$\fvechat_i(t)=R(t)\evechat_i$.  To put this in matrix form we insert a
resolution of the identity,
\begin{equation}
  \fvechat_i(t)= \evechat_j [\evechat_j\cdot(R(t)\evechat_i)]
  = \evechat_j\,R_{ji}(t). 
  \label{fvecevol}
  \end{equation}
From the evolution of the $\fvechat_i$ we obtain that of the $\Fvec_i$
by (\ref{fvechatdef}).  The position as a function of time $\rvec(t)$
is given in terms of these vectors by (\ref{sol}). We omit the details
and focus instead on the asymptotic properties of the solution.

\subsection{Qualitative Features of the Asymptotic Motion}

It is easy to derive some qualitative and semi-quantitative features
of the motion when $t$ is large.  In a uniform magnetic field the
motion has gyrofrequency $\omega_g = e|\Bvec|/mc$ and gyroradius
$r_g=v_\perp/\omega_g$.  In the actual motion for which $\Bvec$ is not
constant, the motion will be qualitatively similar to that in a
uniform field if $\Bvec$ does not change much over the distance $r_g$.
We will call such motion ``adiabatic'' (a word usually defined in a
symplectic context).  For the sphere of monopolium, \eqref{darboux}
gives $\omega_g = h_0 r$ and \eqref{rgeqn} gives $r_g = r_0v_0/h_0 r^2$.
Thus, as the particle moves to larger radii, the gyrofrequency goes to
infinity as $r$ and the gyroradius goes to zero as $1/r^2$.  The
condition that $|\Bvec|$ does not change much in direction over a
distance $r_g$ is $r_g\ll r$, or
\begin{equation}
  \left(\frac{r}{r_0}\right)^3 \gg \frac{v_0}{h_0r_0^2} = 
  \frac{1}{4p^2}.
  \end{equation}
Thus, if $p\ll1$, the motion does not become adiabatic until $r$ has
reached a large multiple of $r_0$, while if $p\gg1$ the motion is
adiabatic for all $t$.  A similar conclusion is reached regarding the
change in the direction of $\Bvec$.

From \eqref{rsqeqn} we have
\begin{equation}
  r(t)= v_0t\sqrt{1 + \frac{r_0^2}{v_0^2 t^2}}\approx v_0t+
  \frac{r_0^2}{2v_0t}+\text{const} + \dots \, .
  \end{equation}
  Then, we can integrate the gyrofrequency $\omega_g(t) = h_0 r(t)$ to find
\begin{equation}
  \int^t dt'\,\omega_g(t') = \frac{h_0v_0t^2}{2} + 
  \frac{h_0r_0^2}{2v_0}\ln t + \text{other terms}.
  \end{equation}
We will see this phase $\phi(t)$ again soon in (\ref{phidef}).
It can be interpreted as the accumulated gyrophase in the asymptotic region.  

\subsection{The Asymptotic Behavior}

To be more quantitative about the asymptotic behavior we invoke the
asymptotic forms of the parabolic cylinder functions (\ref{DnuasympI})
and (\ref{DnuasympII}).  After some work we find the asymptotic form
of the matrix $u$ when $t\to\infty$, including corrections that go as
$1/t$.  We use the frequency $\omega_0=\sqrt{h_0v_0}$ and a
dimensionless time $\tau=\omega_0t$.  We also define
\begin{equation}
  \phi(t)=2p^2\ln\tau + \tau^2/2.
  \label{phidef}
  \end{equation}
Then we find for $t\to+\infty$,
\begin{subequations}
\label{ularget}
\begin{eqnarray}
  u_{11}(t)&=&\frac{e^{-\pi p^2/4}}{2}\left(
    \frac{e^{-i\phi/2}}{A}+
    \frac{p^2e^{i\pi/4}}{\Bbar}\frac{e^{i\phi/2}}{\tau}\right),\\
  u_{12}(t)&=&\frac{e^{-\pi p^2/4}}{2}p\left(
    -\frac{e^{i\pi/4}}{\Bbar}e^{i\phi/2}+
    \frac{e^{-i\phi/2}}{A\tau}\right),\\
  u_{21}(t)&=&\frac{e^{-\pi p^2/4}}{2}p\left(
    \frac{e^{-i\pi/4}}{B}e^{-i\phi/2}
    -\frac{e^{i\phi/2}}{\Abar\tau}\right),\\
  u_{22}(t)&=&\frac{e^{-\pi p^2/4}}{2}\left(
    \frac{e^{i\phi/2}}{\Abar}+\frac{p^2e^{-i\pi/4}}{B}
      \frac{e^{-i\phi/2}}{\tau}\right).
  \end{eqnarray}
\end{subequations}
The forms shown make it easy to check that $u_{21}=-\ubar_{12}$ and
$u_{22}=\ubar_{11}$.  It is also easy  to check that $\det u=1$ to
order $1/\tau$, with the help of the identities (\ref{ABident}).

For negative times we may use the identities
\begin{equation}
  \left(\begin{array}{cc}
      u_{11}(-t) & u_{12}(-t)\\
      u_{21}(-t) & u_{22}(-t)
      \end{array}\right)=
    \left(\begin{array}{cc}
      u_{11}(t) & -u_{12}(t) \\
      -u_{21}(t) & u_{22}(t)
      \end{array}\right),
    \end{equation}
which follow from (\ref{wcol1}) and (\ref{wcol2}).   These are exact
(not just asymptotic), but when $t\to-\infty$ we can use these in
conjunction with (\ref{ularget}) to obtain the asymptotic forms for
$t\to-\infty$.  

\subsection{The $S$-Matrix}

The matrices $u(t)$ and $u(-t)$ do not have a limit as $t\to\infty$
because of the rapidly varying phase $\phi(t)$, but their leading
asymptotic forms (that is, neglecting terms that go as $1/t$) can be
factored,
\begin{equation}
  u(t)=m_+\, q\bigl(\zvechat,\phi(t)\bigr), \qquad
  u(-t)=m_-\,q\bigl(\zvechat,\phi(t)\bigr),
  \end{equation}
in which the $\phi$-dependence appears only in the second factor.
Here $q(\nvechat,\theta)\in SU(2)$ is notation for a spin rotation in
axis-angle form,
\begin{equation}
  q(\nvechat,\theta)=\cos\frac{\theta}{2}-i(\nvechat\cdot\sigmavec)
  \sin\frac{\theta}{2},
  \end{equation}
and the $m$-matrices are given by
\begin{equation}
  m_+ = \frac{e^{-p^2\pi/4}}{2}\left(
    \begin{array}{cc}
      1/A & -pe^{i\pi/4}/\Bbar \\
      pe^{-i\pi/4}/B & 1/\Abar
      \end{array}\right), \qquad
  m_- = \frac{e^{-p^2\pi/4}}{2}\left(
    \begin{array}{cc}
      1/A & pe^{i\pi/4}/\Bbar \\
      -pe^{-i\pi/4}/B & 1/\Abar
      \end{array}\right).
    \label{mmats}
    \end{equation}
The notation $m$ is a mnemonic for M{\o}ller, since these matrices are
analogous to the M{\o}ller wave operator in scattering theory \cite{New}.
We can define $m_\pm(\pm t)$ for any $t>0$ by $u(\pm t)=m_{\pm}(\pm t)
q\bigl(\zvechat,\phi(t)\bigr)$, whereupon the limits
$m_\pm(\pm\infty)$ exist and are given by (\ref{mmats}) (and denoted
simply $m_\pm$).  

The quantities $u(t)$, $m_\pm(t)$ and $q(\nvechat,\phi)$ are elements
of $SU(2)$.  To see the effects of this in three dimensions we write
$R(t)$, $M_\pm(t)$ and $Q(\nvechat,\theta)$ for the corresponding
elements of $SO(3)$.  As for $Q(\nvechat,\theta)$, it is a rotation in
axis-angle form with the same axis and angle as $q(\nvechat,\theta)$.
Then for $t>0$ we have
\begin{equation}
  R(t)=M_+(t)Q\bigl(\zvechat,\phi(t)\bigr), \qquad
  R(-t)=M_-(-t)Q\bigl(\zvechat,\phi(t)\bigr),
  \end{equation}
and (\ref{fvecevol}) implies
\begin{equation}
  \fvechat_i(t)=\evechat_j [M_+(t)Q\bigl(\zvechat,\phi(t)\bigr)]_{ji},
  \qquad\fvechat_i(-t)=\evechat_j [M_-(-t)
    Q\bigl(\zvechat,\phi(t)\bigr)]_{ji}.
    \end{equation}
The vectors $\fvechat_i(\pm t)$ do not approach a limit as
$t\to\infty$, instead $\fvechat_1$ and $\fvechat_2$ spin ever more
rapidly about the direction $\fvechat_3$ as $t$ gets larger.   But we
can strip off this rapid evolution by defining for $t>0$
\begin{equation}
  \avechat_{+i}(t) = \fvechat_j(t) 
  [Q\bigl(\zvechat,-\phi(t)\bigr)]_{ji}, \qquad
  \avechat_{-i}(-t) = \fvechat_j(-t)
  [Q\bigl(\zvechat,-\phi(t)\bigr)]_{ji}.
  \end{equation}
Then the limits $\avechat_{\pm i}(\pm t)$ exist as $t\to\infty$, what
we will call simply $\avechat_{\pm i}$.   These are analogous to the
asymptotic states in the interaction picture in scattering theory (a
version of  the ``in states'' and ``out states'').  

Finally, we define the $S$-matrix by
\begin{equation}
  \avechat_{+i} = \avechat_{-j}\,S_{ji},
  \end{equation}
which implies
\begin{equation}
  S=M_-^{-1}\,M_+.
  \end{equation}
We write $s$ (in lower case) for the corresponding element of $SU(2)$;
it is $m_-^{-1} m_+$.  From (\ref{mmats}) we find
\begin{eqnarray}
  s_{11} &=& s_{22} = e^{-p^2\pi},\\
  s_{21} &=& -\sbar_{12} = \frac{e^{-p^2\pi/2}pe^{-i\pi/4}
    \sqrt{2\pi}}{\Gamma(1-\nu)}
  =e^{i(\eta-\pi/4)}\,\sqrt{1-e^{-2\pi p^2}},
  \end{eqnarray}
where $\eta=\arg\Gamma(1+\nu)$.  By comparing this with (\ref{uRproj})
we see that $s$ is a rotation by an angle $\theta$ about an axis
$\bvechat$ in the $x$-$y$ plane, $s=q(\bvechat,\theta)$, where
\begin{equation}
  \cos\frac{\theta}{2}=e^{-p^2\pi}, \qquad
  \bvechat = \xvechat\cos\alpha+\yvechat\sin\alpha,
  \label{scattangle}
  \end{equation}
where $\alpha=\eta+\pi/4$.  

Thus the asymptotic frame vectors are related by
\begin{equation}
  \avechat_{+i}=\avechat_{-j}\,Q(\bvechat,\theta)_{ji}.
  \end{equation}
The particle comes in asymptotically along a radial line and goes out
asymptotically along a different radial line, and the scattering angle
is $\theta$, given by (\ref{scattangle}).  However the azimuthal angle
of the scattering, measured with respect to the incoming direction,
depends ever more sensitively on the azimuthal angle of the angular
component of the velocity in the remote past.   The asymptotic states
are not free, and a cross section in the usual sense does not exist.

\subsection{Intuitive Picture of the Trajectory}

When the particle is far from the origin, the magnetic field is strong.
The particle's motion can be decomposed into radial motion (aligned with 
the magnetic field) and rapid gyration about the magnetic field.  Far
from the origin, the motion is nearly in the radial direction.
The more interesting dynamics occurs when the particle ``scatters'' off the 
weaker field near the origin. The behavior depends on the magnitude 
of the dimensionless parameter $p$.

\begin{figure}
\includegraphics[scale=0.5]{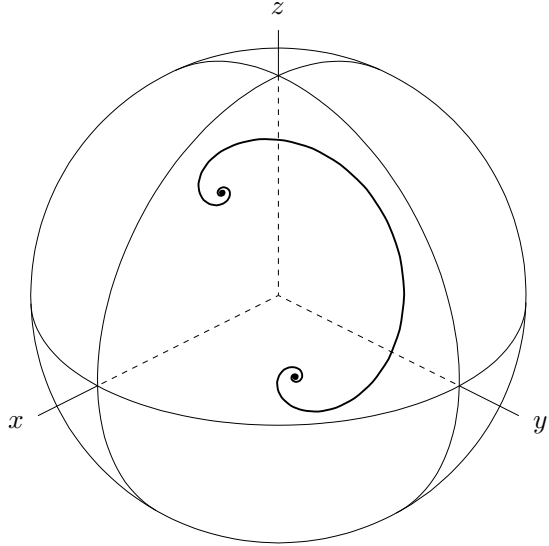}%
\caption{\label{orbit}An orbit projected onto the unit sphere, with
  $p=0.553$, corresponding to a scattering angle of
  $\theta=135^\circ$.  The orbit has been rotated to appear in the
  principal octant of the sphere.   The initial point at $t=0$ is the
  symmetry point.   The asymptotic spirals as $t\to\pm\infty$ are
  clearly seen.}%
\end{figure}

For large $p$, the particle never gets close enough to the origin to
experience a weak magnetic field. The particle remains ``adiabatic''
for all time and must stay close to a magnetic field line in one
radial direction.  The particle approaches the origin, then reflects
and returns the way it came.  The scattering angle is close to $\pi$.

For small $p$, the particle does get close enough to the origin to
experience a weak magnetic field. Near the origin, the motion is
approximately force free, so the particle goes past the origin and
continues on the other side.  The scattering angle is close to zero.

Figure~\ref{orbit} illustrates an orbit with initial conditions
$r_0=v_0=1$ and $p=0.553$, which, according to (\ref{scattangle}),
corresponds to a scattering angle of $\theta=135^\circ$.  For purposes
of illustration it is convenient to project the orbit onto the unit
sphere, as has been done in the figure.  The asymptotic spirals as
$t\to\pm\infty$ can be seen, corresponding to the incoming and
outgoing directions.  The point of symmetry on the orbit is $t=0$, the
point of minimum approach where $r=r_0$.  The orbit illustrated is not
the fundamental solution, but rather for the purpose of the
presentation it has been rotated to place the orbit entirely within the
principal octant of the sphere.   More extensive numerical work
confirms the formula (\ref{scattangle}) for the scattering angle.

\subsection{Cones and Geodesics}
\label{conesgeodesics}

The case $\rho(\rvec)=g_0\delta^3(\rvec)$, $\Bvec=g_0\rvec/r^3$ is the
point monopole at the origin, for which the solution is well known.
In this case Bakas and L\"ust \cite{BL} have provided an illustration
of the orbit on the cone.  We offer another way of visualizing this
solution.

Poincar\'e \cite{Po1896} pointed out that the orbit is a geodesic on
the cone.   This is because vectors $\rvec$ and $\vvec$, which are
linearly independent and which span the tangent plane to the cone, are
orthogonal to the force vector.   Therefore the magnetic force is a
force of constraint on the cone, and the motion is a geodesic.

Since however a cone is isometric to a plane, the orbit may be
visualized by drawing a straight line on a piece of paper, which is
then rolled into a cone.  One can clearly see the effective repulsion
of the particle from regions of high magnetic field strength (the
mirror effect).  One line suffices to visualize a continuum of
possible orbits that is, for various initial conditions, as the paper
is rolled into a cone in different ways.

It turns out that Poincar\'e's construction generalizes to arbitrary,
spherically symmetric distributions of monopoles.  To do this we
define a generalized cone as the union of a smooth family of radial
half-lines, emanating from the origin.  The cone that applies to
motion in a spherically symmetric distribution of monopoles is the one
swept out by the half-line joining the origin with the particle
position $\rvec(t)$ and extending out to infinity, as $t$ goes from
$-\infty$ to $+\infty$.  Such a cone is conveniently visualized by its
intersection with the unit sphere, which is a curve.  This curve may
self-intersect (although not in our examples), and as $t$ goes from
$-\infty$ to $+\infty$ parts of it may be retraced.

In the case of a point monopole, the cone in this sense is the same as
the usual cone, and the curve on the unit sphere is a small circle, or
an arc thereof; for a free particle ($\Bvec=0$) the curve is an arc of
a great circle; and for the uniform distribution of monopolium an
example of the curve on the unit sphere is given in Fig.~\ref{orbit}.

The cone in this sense is locally isometric to the Euclean plane.  Let
$s$ be the arc length of the curve on the unit sphere.  Then an
increment of arc length on the unit sphere $ds=d\theta$ is also an
increment of angle, and the sector of the cone defined by $d\theta$ is
isometric to the corresponding sector of the Euclidean plane where
$\theta$ is the usual polar coordinate.  More formally, if $\theta$ is
a length or accumulated angle coordinate along the curve on the unit
sphere, then $(r,\theta)$ are coordinates of points on the cone, and
the metric on the cone, as inherited from the Euclidean metric in
$\Reals^3$, is $dr^2+r^2d\theta^2$, the same as the metric in
Euclidean $\Reals^2$ in polar coordinates.  Poincar\'e's argument
applies to the cone as we have defined it, and the orbits are
geodesics, the images of straight lines on $\Reals^2$ under the
mapping in which points are identified by their $(r,\theta)$
coordinates.
\section{Conclusions}

In this article we have analyzed the motion of a charged particle in
the magnetic field produced by a spherically symmetric distribution of
magnetic monopoles.  We have found some general features of such
solutions, that is, for any spherically symmetric distribution of
monopoles, including the fact that the equations of motion can be
converted into a linear system and the orbit is a geodesic on a
generalized cone.   In the special case of a uniform, spherically
symmetric distribution of monopoles we have given a complete solution
for the orbits, including an $S$-matrix that connects asymptotic
states and an explicit formula for the scattering angle.   These
results enlarge the repertoire of systems of distributions of magnetic
monopoles for which the orbits of charged particles are known.

\begin{acknowledgments}
A portion of this research was carried out at MSRI, supported in part by NSF grant DMS-0441170, while the authors participated in  the program {\it Hamiltonian systems, from topology to applications through analysis}.  JH  and PJM were also supported by the DOE Office of Fusion Energy Sciences under DE-FG02-04ER-54742.
\end{acknowledgments}

\appendix

\section{Parabolic Cylinder Functions}
\label{pcfuns}

We have found that the best reference for parabolic cylinder functions
for our purposes is Magnus and Oberhettinger \cite{MO}.  The edition
cited has a small and obvious error in the asymptotic forms, which has
been corrected in (\ref{DnuasympI}) and (\ref{DnuasympII}); earlier
editions have more serious errors.  

Parabolic cylinder functions, denoted $D_\nu(z)$, are entire analytic
functions of $z$.  In this appendix $z$ and $\nu$ are arbitrary
complex numbers, while in the main body of the paper $z$ and $\nu$
have the specific values (\ref{zdef}) and (\ref{nudef}).  The
parabolic cylinder functions satisfy the differential equation,
\begin{equation}
  \frac{d^2f(z)}{dz^2} + \left(\nu+\frac{1}{2} -\frac{z^2}{4}
    \right)f(z)=0,
    \label{pcode}
  \end{equation}
of which the functions $D_\nu(\pm z)$ and $D_{-\nu-1}(\pm iz)$ are
solutions.  These have the linear dependencies,
\begin{eqnarray}
  D_\nu(z) &=& e^{-i\nu\pi}\,D_\nu(-z) +
  \frac{\sqrt{2\pi}}{\Gamma(-\nu)}\,e^{-i(\nu+1)\pi/2}\,
  D_{-\nu-1}(iz)\nonumber\\
  &=& e^{i\nu\pi}\,D_\nu(-z)+
  \frac{\sqrt{2\pi}}{\Gamma(-\nu)}\,e^{i(\nu+1)\pi/2}\,
  D_{-\nu-1}(-iz).
  \label{lindeps}
  \end{eqnarray}
These functions satisfy the recursion relations,
\begin{subequations}
\label{recursionrels}
\begin{eqnarray}
    D'_\nu(z)+\frac{z}{2}D_\nu(z)-\nu D_{\nu-1}(z)&=&0,
    \label{recursionrelsa}\\
    D'_\nu(z)-\frac{z}{2}D_\nu(z)+D_{\nu+1}(z)&=&0,
    \label{recursionrelsb}
    \end{eqnarray}
\end{subequations}
and the initial conditions,
\begin{equation}
  D_\nu(0)=\frac{\sqrt{\pi}\,2^{\nu/2}}{\Gamma\bigl((1-\nu)/2\bigr)},
  \qquad
  D'_\nu(0)=-\frac{\sqrt{\pi}\,2^{(\nu+1)/2}}{\Gamma(-\nu/2)}.
  \label{pcinitcond}
  \end{equation}

The parabolic cylinder functions have the integral representation,
\begin{equation}
  D_\nu(z) = \frac{e^{-i\nu\pi/2}}{\sqrt{2\pi}}
  \int_C dt\, t^\nu\, e^{-t^2/2+itz+z^2/4},
  \label{Dnuintegralrepn}
  \end{equation}
where the contour $C$ runs just above the real axis and
where $t^\nu$ has a branch cut just below the positive real axis.
This is not the same as the integral representations quoted in
\cite{MO}, but is convenient because it applies  to all
$\nu\in\Complexes$.   Equation~(\ref{Dnuintegralrepn}) can be proved
by showing that $D_\nu(z)$, so defined, satifies the differential
equation (\ref{pcode}) and the initial conditions (\ref{pcinitcond}).
 
When $|z|\gg 1, |\nu|$, the integral representation
(\ref{Dnuintegralrepn}) leads via the method of steepest descent to
the asymptotic forms,
\begin{equation}
  D_\nu(z)\approx e^{-z^2/4}\,z^\nu\,\left[
    1-\frac{\nu(\nu-1)}{1!\cdot 2z^2}
    +\frac{\nu(\nu-1)(\nu-2)(\nu-3)}{2!\cdot 2^2z^4} -\ldots\right],
  \label{DnuasympI}
  \end{equation}
for $-\pi/2 < \arg z < \pi/2$, and
\begin{eqnarray}
  D_\nu(z) &\approx& \hbox{\rm Series (\ref{DnuasympI})}
  -\frac{\sqrt{2\pi}}{\Gamma(-\nu)}\,
  e^{i\nu\pi}\,e^{z^2/4}\,z^{-\nu-1}\nonumber\\
   &\times&\left[
    1+\frac{(\nu+1)(\nu+2)}{1!\cdot 2z^2}
    +\frac{(\nu+1)(\nu+2)(\nu+3)(\nu+4)}{2!\cdot 2^2 z^4}
    +\ldots\right],
  \label{DnuasympII}
  \end{eqnarray}
for $\pi/2 < \arg z < \pi$.  The asymptotic form changes at $\arg
z=\pi/2$ because of a bifurcation in the steepest descent contour; see
Dingle \cite{Ding} for details.  For the purposes of this paper we
need only $\arg z=-\pi/4$ and $3\pi/4$.


\begin{thebibliography}{9}

\bibitem{BL}I. Bakas and D. L\"ust, J. High Energy Phys. {\bf 01},
  171(2014).

\bibitem{PJM82} P. J. Morrison, AIP Conf. Proc. {\bf 88}, 13(1982).

\bibitem{HM20} J. M. Heninger and P. J. Morrison, Phys. Lett. A {\bf 384},
126101(2020).

\bibitem{Cetal}J. F. Cari\~nena, J. M. Gracia-Bond\'\i a, F. Lizzi,
G. Marmo and P. Vitale, Phys. Lett. A {\bf 374}, 3614(2010).

\bibitem{Sol}M. A. Soloviev, J. Phys. A {\bf 51}, 095205(2018).

\bibitem{KS}V. G. Kupriyanov and R. J. Szabo, Phys. Rev. D {\bf 98},
  045005(2018).

\bibitem{LSW}M. Lainz, C. Sard\'on and A. Weinstein, Phys. Rev. D {\bf
    100}, 105016(2019).

\bibitem{Mo80}P. J. Morrison, Phys. Lett. A {\bf 80}, 383(1980).

\bibitem{WM81}A. Weinstein and P. J. Morrison, Phys. Lett. A {\bf 86},
  235(1981).

\bibitem{MW82}J. E. Marsden and A. Weinstein, Physics {\bf 4D}, 394(1982).

\bibitem{Li83}R. G. Littlejohn, J. Plasma Phys. {\bf 29}, 111(1983).

\bibitem{Po1896} H. Poincar\'e, Compt. Rend. {\bf 123}, 530(1896).

\bibitem{Ar78} V. I. Arnold, {\it Mathematical Methods of Classical Mechanics}
(Springer, New York, 1978).

\bibitem{Blaschke} W.  Blaschke,  \textit{Vorlesungen \"Uber 
Differentialgeometrie I} (Springer, Berlin, 1930).

\bibitem{FL} W. G. Flynn and R. G. Littlejohn, Ann. Phys. {\bf 234},
  334(1994).

\bibitem{YCdV} Y. Colin de Verdi\`ere, Annales de l'Institut Fourier
  {\bf 53}, 1023(2003).

\bibitem{TK} E. R. Tracy and A. N. Kaufman, Phys. Rev. Lett. {\bf 91},
  130402(2003).

\bibitem{MO} W. Magnus and F. Oberhettinger, \textit{Formulas and
    Theorems for the Special Functions of Mathematical Physics}
  (Chelsea Publishing, New York, 1949).

\bibitem{New} R. G. Newton, \textit{Scattering Theory of Waves and
    Particles, 2nd ed.} (Springer-Verlag, New York, 1982).

\bibitem{Ding} R. B. Dingle, \textit{Asymptotic Expansions: Their
    Derivation and Interpretation} (Academic Press, London, 1973). 

\end{thebibliography}

\end{document}